\documentstyle[12pt]{article} 
\newcommand{\dd}{\mbox{\rm d}}
\newcommand{\nnn}{\noindent}
\newcommand{\be}{\begin{equation}}
\newcommand{\ee}{\end{equation}}
\newcommand{\bi}{\bibitem}

\begin{document}
\rightline{IJS-TP-95/10}

\vspace{1.5cm}

\begin{center} {\bf \large On the Resolution of Time Problem in Quantum
Gravity Induced from Unconstrained Membranes}

\vspace{.5cm}

Matej Pav\v si\v c

\vspace{.3cm}

Jo\v zef Stefan Institute, University of Ljubljana, Ljubljana, Slovenia

\vspace{1.5cm}

ABSTRACT

\end{center}

The relativistic theory of unconstrained $p$-dimensional membranes
($p$-branes) is further developed and then applied to the embedding
model of induced gravity. Space-time is considered as a 4-dimensional
unconstrained membrane evolving in an $N$-dimensional embedding space.
The parameter of evolution or the evolution time $\tau$ is a distinct
concept from the coordinate time $t = x^0$. Quantization of the theory is
also discussed. A covariant functional Schr\" odinger equations has
a solution for the wave functional such that it is sharply localized in
a certain subspace $P$ of space-time, and much less sharply localized
(though still localized) outside $P$. With the passage of evolution the
region $P$ moves forward in space-time. Such a solution we interpret
as incorporating two seemingly contradictory observations:
(i) experiments clearly indicate that space-time is a continuum in
which events are existing; (ii) not the whole 4-dimensional space-time,
but only a 3-dimensional section which moves forward in time
is accessible to our immediate
experience. The notorious problem of time is thus resolved in our
approach to quantum gravity.
Finally we include sources into our unconstrained embedding model.
Possible sources are unconstrained worldlines which are free from the
well known problem concerning the Maxwell fields generated by charged
unconstrained point particles.

\vspace{1cm}

Short title: Resolution of Time Problem in Quantum Gravity
\newpage

{\bf 1.  Introduction}

Since the pioneering works of Sakharov \cite{1} and Addler \cite{2}
there has been increasing
interest in various models of the induced gravity \cite{3}.
A particularly interesting
and promising seems to be the model in which spacetime is a 4-dimensional
manifold
(a "spacetime sheet") $V_4$ embedded in an $N$-dimensional space $V_N$
\cite{4}-\cite{6a}.  The
dynamical variables are the embedding functions ${\eta}^a(x)$ which determine
positions (coordinates) of points on $V_4$ with respect to $V_N$.  The action
is
a straightforward generalization \cite{6},\cite{6a}
of the Dirac-Nambu-Goto action. The latter can be
written in an equivalent form in which there appears the induced metric $g_{\mu
\nu}(x)$ and ${\eta}^a(x)$ as variables which have to be varied independently.
Quantization of such action enables one to express {\it an effective action} as
a functional of $g_{\mu \nu}(x)$.  The effective action is obtained in the
Feynman path integral in which we functionally integrate over the embedding
functions ${\eta}^a(x)$ of $V_4$, so that what remains is a functional
dependence on $g_{\mu \nu}(x)$.  Such an effective action containes the Ricci
curvature scalar $R$ and its higher orders \cite{3}.
This theory was discussed more detailly in a previous work \cite{6a}.

In the present paper we are going to generalize the above approach.  The main
problem with any reparametrization invariant theory is the presence of
constraints relating the dynamical variables.  Therefore there exist
equivalence classes of functions ${\eta}^a (x)$ - related by reparametrizations
of the coordinates $x^{\mu}$ - such that each member of an equivalence class
represents the same spacetime sheet $V_4$.  This must be taken into account in
the quantized theory, e.g.  when performing, for instance a functional
integration over ${\eta}^a (x)$.  Though elegant solution to such problems were
found in string theories \cite{7}, the technical difficulties accumulate
in the case of a $p$-dimensional membrane ($p$-brane)
with $p$ greater than 2 \cite{7a}.

We first discuss the possibility of removing constraints from a
membrane ($p$-brane) theory.  Such a generalized theory possesses additional
degrees of freedom and contains the usual $p$-branes of the Dirac-Nambu-Goto
type
as a special case.  It is an extension, from a point-particle to a
$p$-dimensional membrane, of a theory which treats a relativistic particle
without constraint, so that all coordinates $x^{\mu}$ and the conjugate momenta
$p_{\mu}$ are independent dynamical variables which evolve along the invariant
evolution parameter $\tau$ \cite{8}-\cite{9a}.
A membrane is then considered as a continuum of
such point particles and has no constraints.  It was shown \cite{10},\cite{10a}
that the extra
degrees of freedom are related to variable stress and fluid velocity on the
membrane, which is therefor, in general, a "{\it wiggly membrane"}.  Then we
apply the concept of a relativistic membrane without constraints to the
embedding
model of induced gravity in which the whole spacetime is considered as a
membrane in a flat embedding space.

In Sec. 2 we develop the theory of an {\it unconstrained} relativistic
$p$-brane
(also called simply membrane, with understanding that its dimension $p$ is
arbitrary), denoted ${\cal V}_p$ (in constrast to a constrained membrane
$V_p$).
To facilitate the introduction of our concepts we use the usual
notation, where variables $X^{\mu}, \; \mu = 0,1,2,...,D-1$ , represent
coordinates of a membrane living in a $D$-dimensional spacetime, and ${\xi}^a ,
\; a = 0,1,2,...,d-1$ , are parameters of a worldsheet $V_d$ swept by membrane
(with $p = d - 1$).

In Sec. 3 we apply the theory of Sec.2 to the concept of an $(n-1)$-
dimensional
simultaneity surface ${\cal V}_{n-1}$ (analogous to a $p$-brane of Sec.2)
moving in an
$N$-dimensional embedding space $V_N$ and thus sweeping a space-time sheet
$V_n$
(analogous to the worldsheet $V_d$ of Sec.2).  Notation is here
changed, and in some sense reversed:  ${\eta}^a (x), a=0,1,...,N-1$ are
positions of a spacetime surface (called also sheet) $V_n$ in the embedding
space $V_N$, and $x^{\mu}, \; \; \mu =0,1,...,n-1$ are parameters (coordinates)
on $V_n$.

In Sec. 4. we consider the theory in which
the whole space-time is an $n$-dimensional\footnote
{Usually $n=4$, but if we wish
to consider a Kaluza-Klein like theory, then $n > 4$}
unconstrained membrane ${\cal V}_n$
analogous to a $p$-brane ${\cal V}_p$
of Sec.2.  The theory allows for {\it motion} of
${\cal V}_n$ in the embedding space $V_N$.  When considering the quantized
theory it
turns out that a particular wave packet functional exists such that:

\hspace{.5cm} (i) it approximately represents evolution of
a simultaneity surface ${\cal V}_{n-1}$ (also denoted
${\cal V}_{\Sigma}$), and

\hspace{.4cm} (ii) all possible space-time membranes ${\cal V}_n$ composing
the wave packet
are localized near an average space-time membrane ${\cal V}_n^{(c)} $
which corresponds to a classical space-time unconstrained membrane.

This approach gives both:  the evolution of a state (to which classically there
corresponds the progression of time slice) and a fixed spacetime as the
expectation value.  The notorious problem of time, as it occurs in a
reparametrization invariant theory (for instance in general relativity), does
not exist in our approach.

\vspace{1cm}

{\bf 2.  Relativistic membranes without constraints

2.1.  A reformulation of the conventional p-brane action}

\vspace{.3cm}

Relativistic $p$-dimensional constrained membranes \cite{7a},
including strings ($p=1$) \cite{7}
and point particles ($p=0$), are commonly described by an action which is
invariant under reparametrizations of of coordinates ${\xi}^a, a=0,1,2,...,p$,
of the $d=p+1$ dimensional worldsheet $V_d$ swept by a $p$-dimensional
membrane.
Consequently, the dynamical variables $X^{\mu}, \mu =0,1,2,...,D$ and the
corresponding momenta are subjected to $d$ primary constraints; not all
$X^{\mu}$ are independent, there are $d$ relations among them.

A suitable form of the action \cite{13}
(equivalent to the Dirac-Nambu-Goto action \cite{14}) is
\be
   I[X^{\mu},{\gamma}^{ab}] = {{\kappa} \over 2} \int {\dd}^d \xi
   \sqrt{| \gamma |} \left ( {\gamma}^{ab} {\partial}_a X^{\mu} {\partial}_b
   X^{\nu} g_{\mu \nu} + 2 - d \right )
\label{1}
\ee

\nnn where $X^{\mu}$ and ${\gamma}^{ab}$ are to be varied independently.
Variation
of ${\gamma}^{ab}$ gives the expression for the induced metric ${\gamma}_{ab} =
{\partial}_a X^{\mu} {\partial}_b X_{\mu}$ on $V_d$.  The Lagrange multipliers
${\gamma}^{ab}$
are not all independent:  there are $d(d+1)/2$ components of ${\gamma}^{ab}$ ,
while there are only $d$ constraints.

In order to separate $d$ independent Lagrange multipliers
we perform an ADM-like \cite{15} decomposition of $V_d$ such that \cite{16}
\be {\gamma}^{ab} = {{n^a n^b} \over {n^2}} + {\bar \gamma}^{ab}
     \label{3} \ee
where $n^a$ is the normal vector field to a chosen
hypersurface $\Sigma$ and ${\bar \gamma}^{ab}$ the projection tensor which
projects an arbitrary vector into $\Sigma$.  For instance, ${\bar
\gamma}^{ab}$ projects a derivative ${\partial}_a X^{\mu}$ into the tangent
derivative:
\be {\bar \partial}_a X^{\mu} = {{\bar \gamma}_a}^{\, b}
  {\partial}_b X^{\mu} = {\partial}_a
  X^{\mu} - n_a {\partial} X^{\mu}  \label{3a} \ee

\nnn where ${\partial} X^{\mu} \equiv n^b {\partial}_b X^{\mu}/n^2$ is the
normal
derivative.

Let us take such a class of coordinates system in which covariant components of
normal vectors are $n_a = (1,0,0,...,0)$.  Then we have
\be
  {\gamma}^{00} = n^0 = n^a n_a \; \; , \quad {\gamma}^{0i} = n^i \; \; \quad
  {\gamma}^{ij} = {\bar \gamma}_{ij} + n^i n^j/n^0
\label{3b}
\ee
\be {\gamma}_{00} = {1 \over {n^0}} + {\bar \gamma}_{ij} {{n^i n^j} \over
   {(n^0)^2}} \; ,, \quad {\gamma}_{0i} = - {\bar \gamma}_{ij} n^j/n^0 \; ,
   \quad {\gamma}_{ij} = {\bar \gamma}_{ij} \; , \quad i,j = 1,2,..., p
\label{3c}
\ee

\nnn The decomposition (\ref{3a}) then becomes
\be {\partial}_0 X^{\mu} = {\partial} X^{\mu} + {\bar \partial}_0 X^{\mu}
    \label{3d} \ee
\be {\partial}_i X^{\mu} = {\bar \partial}_i X^{\mu}   \label{3e} \ee

\nnn where
\be {\dot X}^{\mu} \equiv {\partial}_0 X^{\mu} \equiv {{{\partial} X^{\mu}}
\over
  {\partial {\xi}^0}} \label{3f} \ee
\be {\partial} X^{\mu} = {\dot X}^{\mu} + {{n^i {\partial}_i X^{\mu}} \over
{n^0}} \;
   \; \quad {\partial}_i X^{\mu} \equiv {{{\partial} X^{\mu}} \over {\partial
   {\xi}^i}}
   \label{3g} \ee

\nnn As $d$ independent Lagrange multipliers can be taken $n^a = (n^0,n^i)$.
We can
now rewrite our action in terms of $n^0$ and $n^i$.  We insert (\ref{3b}) into
(\ref{1}) and take into account that
\be \vert \gamma \vert = {{\bar \gamma} \over {n^0}} \label{3h} \ee

\nnn where $\gamma = {\mbox{\rm det}}\, {\gamma}_{ab}$ is the determinant of
the
worldsheet metric and ${\bar \gamma} = \mbox{\rm det} {\bar \gamma}_{ij}$ the
determinant of the metric ${\bar \gamma}_{ij} = {\gamma}_{ij}$ on the
hypersurface $\Sigma$.

After using (\ref{3}-\ref{3h}) our action (\ref{1}) becomes a functional
$I[X^{\mu},n^a,{\bar \gamma}^{ij}]$ of $X^{\mu}$ and the Lagrange multipliers
$n^a, \; {\bar \gamma}_{ij}$.  Variation of $I[X^{\mu},n^a,{\bar \gamma}^{ij}]$
with respect to ${\bar \gamma}^{ij}$ gives the expression for the induced
metric on the surface $\Sigma$
\be {\bar \gamma}_{ij} = {\partial}_i X^{\mu} {\partial}_j X_{\mu} \; \; \quad
{\bar
   \gamma}^{ij} {\bar \gamma}_{ij} = d - 1  \label{3i} \ee

\nnn Using the latter expression (\ref{3i}) we can eliminate ${\bar
\gamma}^{ij}$ from $I[X^{\mu},n^a,{\bar \gamma}^{ij}]$ and we obtain \cite{16},
\cite{10} a functional
of $X^{\mu}$ and $d$ independent Lagrange multipliers $n^a = (n^0,n^i)$
\be I[X^{\mu},n^a] = {{\kappa} \over 2} \int {\mbox{\rm d}} \tau
   {\mbox{\rm d}}^p
    \sigma \sqrt{\bar f} \left ( {{{\partial} X^{\mu} {\partial} X_{\mu}}
    \over {\lambda}} + \lambda \right ) \; \; , \quad \; \lambda \equiv
    {1 \over \sqrt{n^0}}
\label{2}
\ee

\nnn where ${\partial} X^{\mu}$ is given by (\ref{3g}) and ${\bar f} \equiv
\mbox{\rm det} {\bar f}_{ij}, \; {\bar f}_{ij} \equiv {\partial}_i X^{\mu}
{\partial}_j X_{\mu}$.  In eq.(\ref{2}) the coordinates are split according to
${\xi}^a = ({\xi}^0,{\xi}^i) \equiv (\tau , {\sigma}^i)$ and the volume element
written as ${\mbox{\rm d}}^d \xi = {\dd} \tau {\mbox{\rm d}}^p \sigma$. Instead
of $n^0$ a new symbol $\lambda \equiv 1/\sqrt{n^0}$ is introduced.

So we have arrived at an action which looks like the well known Howe-Tucker
action for a point particle, apart from the integration over coordinates
${\sigma}^i$ of a space-like hypersurface $\Sigma$.  Indeed, eq.(\ref{2}) is an
action of a continuous collection of point particles:  it is a functional of a
bundle $X^{\mu} (\tau, {\sigma}^i)$ of worldlines.  Individual worldlines are
distinguished by the values of parameters ${\sigma}^i$.

The equations of motion for the variables $X^{\mu}$ derived from (\ref{2}) are
exactly the equations of a minimal surface (as derived directly from
(\ref{1})); the equations of "motion" for $n^a$ are the worldsheet
constraints \cite{10},\cite{16}.

\vspace{1cm}

{\bf 2.2 A generalized} $p$-{\bf brane action}

\vspace{.3cm}

So far we had just a suitable reformulation of the well known Dirac-Nambu-Goto
theory of minimal surfaces.  Now we shall do a crucial step:  let us fix $n^a =
(n^0,\, n^i)$ in eq.(\ref{2}) so that $n^a$ are no longer Lagrange multipliers,
but given functions of $\tau$ and $\sigma$.  In the conventional approaches
such
a fixing is interpreted as {\it gauge fixing} and the action (\ref{2}) with
fixed $n^a$ contains, besides the physical ones, also the unphysical degrees of
freedom which must be compensated by an additional term, the so called gauge
fixing term.

My aim is to go beyond the conventional theory. As an alternative to
the conventional action I proposed \cite{10},\cite{10a}
the following new action \footnote{The old action
(\ref{2}) served only as a guidance for introducing the new action
(\ref{5}).
While the action (\ref{2}) is equivalent to the conventional action (1),
the new
action (\ref{5}) is not equivalent to (\ref{1}).  The step from (\ref{2}) to
(\ref{5}) by fixing $n^0 \ne 0$ and $n^i = 0$ brings
a new physical content into the
membrane's theory.  In this paper we consider a particular fixing of $n^i$,
namely $n^i = 0$.  A more general choice of $n^i$ is also possible \cite{16}.}:
\be I[X^{\mu}] = {{\kappa} \over 2} \int {\mbox{\rm d}} \tau {\mbox{\rm d}}^p
	\sigma \sqrt{\bar f} \left ( {{{\dot X}^{\mu} {\dot
	X}_{\mu}} \over {\Lambda}} + \Lambda \right )
\label{5}
\ee

\nnn Our action (\ref{5}) is not
invariant with respect to reparametrizations of $\tau$.  The latter is not an
arbitrary parameter, but it is a fixed parameter of the true dynamical
evolution.
Consequently, all $D$ variables $X^{\mu}(\tau,\, {\sigma}^i)$ are independent
physical dynamical degrees of freedom.  There are no constraint among $X^{\mu}$
and the corresponding canonically conjugate momenta
\be p_{\mu} = {{\partial L} \over {\partial {\dot X}^{\mu}}} = \sqrt{\bar f} \,
    {{{\dot X}^{\mu}} \over {\Lambda}} \label{5a} \ee

\nnn We do not need to introduce ghosts in (\ref{5}), since we consider the
extra degrees of freedom as physical ones and not merely related to choice of
gauge.

In the action (\ref{5}) $\Lambda$ is a given function of $\tau$ and
${\sigma}^i$.  Different choices of $\Lambda(\tau,\sigma)$ correspond to
physically different dynamical systems.  The new symbol $\Lambda$ is necessary
in (\ref{5}) in order to distinguish the new action (in which $\Lambda$ is not
a Lagrange multiplier) from the old action (\ref{2}) (in which $\lambda$ is a
Lagrange multipier).

The equations of motion derived from (\ref{5}) are no longer those of a minimal
surface.  In Refs. \cite{10},\cite{10a}
it was shown that such equations of motion are equivalent
to the equations of a so called {\it wiggly membrane} \cite{19}
with a particular equation
of state.  The extra degrees of freedom of our unconstrained membrane
${\cal V}_p$ are
related to the variable tension and fluid velocity of a wiggly membrane.

Our action (\ref{5}) is invariant with respect to reparametrizations
${\sigma}^i
\rightarrow {\sigma}^{\prime i} = {\sigma}^{\prime i} (\sigma)$ of the
parameters on a membrane ${\cal V}_p$.  But, since this invariance
does not involve
the evolution parameter $\tau$, it does not imply constraints among the
dynamical variables $X^{\mu}(\tau,\sigma)$ (and
the corresponding momenta $p_{\mu}(\tau,\sigma)$).

We can introduce the auxiliary variables ${\bar \gamma}^{ij}$ and write an
action which is equivalent to (\ref{5}):
\be I[X^{\mu},{\bar \gamma}^{ij}] = {{\kappa} \over 4} \int \dd \tau {\dd}^p
    \sigma \left ( \sqrt{|{\bar \gamma}|} {\bar \gamma}^{ij} {\partial}_i
X^{\mu}
    {\partial}_j X_{\mu} + 2 - p) \right )
    \left ( {{{\dot X}^2} \over {\Lambda}} + \Lambda
    \right )  \label{6} \ee

\nnn Variation of (\ref{6}) with respect to ${\bar \gamma}^{ij}$ gives the
relation

\be {\bar \gamma}_{ij} = {\partial}_i X^{\mu} {\partial}_j X_{\mu}  \label{7}
\ee

\nnn which tells that ${\bar \gamma}^{ij}$ is the induced metric on the
$p$-dimensional surface ${\cal V}_p$ in $D$-dimensional embedding space.
Eq.(\ref{7}) does not express constraints among $X^{\mu}$, it is merely the
definition equation for the auxiliary variables ${\bar \gamma}^{ij}$ .

The initial conditions for our dynamical system described by the action
(\ref{5}) or (\ref{6}) are given by
\be X^{\mu}(\tau=0,\, \sigma) \label{8} \ee \be {\dot X}^{\mu}(\tau=0, \,
   \sigma) \label{9} \ee

\nnn Equation (\ref{8}) determines a $p$-dimensional unconstrained membrane
${\cal V}_p$ at the
initial value of the parameter $\tau$.  Equation (\ref{9}) determines a
field of velocities at $\tau = 0$.  A solution of the dynamical equations
derived from (\ref{6}) determines the membrane ${\cal V}_p$
at various values of parameter $\tau$; in other words, it determines
membrane's motion in space-time.  When ${\cal V}_p$
{\it moves} it sweeps a $d$-dimensional surface $V_d$.  An initial
${\cal V}_p$
is given arbitrarily, and its parametric equation is $x^{\mu} =
X^{\mu}({\sigma}^i) \; , \; \mu = 0,1,..., D-1\, ; \; \; i,j =
0,1,...,d-1$ .  Once
${\cal V}_p$ is given, we can calculate the tangential derivatives
${\partial}_i
X^{\mu}$ and the induced metric (\ref{7}).  This illustrates that none of the
tangential derivatives ${\partial}_i X^{\mu}$ (and consequently the induced
metric
${\bar \gamma}_{ij}$) is given independently as initial condition, and
therefore eq.(\ref{7}) indeed does not imply any constraint among the dynamical
variables $X^{\mu}(\tau,\sigma)$.  Though in our theory a
local gauge group is present (invariance of the action (\ref{5}) under
reparametrizations of ${\sigma}^i$), yet
solving the equations of motion derived from
the action (\ref{5}) with a given set of initial data (\ref{8}),(\ref{9})
constitutes a well-posed Cauchy problem. This is true, because the gauge
group does not involve the evolution parameter $\tau$.\footnote
{Analogous situation occurs in the description of
non-relativistic motion of a usual 1-dimensional string or 2-dimensional
membrane in 3-dimensional space, with the ordinary time $t$ as evolution
parameter.  The fact that one can arbitrarily parametrize string or membrane
does not imply dynamical constraints in such a non-relativistic motion.}

Let me try to further clarify this point. First, we assume that points
of the membrane ${\cal V}_p$ are physically distinguishable. They can be
marked and later, after the passage of evolution again identified. Next,
we give "house numbers" to the points on ${\cal V}_p$ , that is, we choose
a parametrization ${\sigma}^i$. The choice is arbitrary, and the
theory is invariant under reparametrizations ${\sigma}^i \rightarrow
{\sigma '}^i = {\sigma '}^i ({\sigma}^i)$. Suppose we have chosen an initial
membrane and fixed a parametrization (i.e. choice of coordinates ${\sigma}^i$)
on it, so that $X^{\mu} (0,\sigma)$ is given. The
dynamical equations of motion (straightforwardly derived from the action
(\ref{5})
and the initial data (\ref{8}),(\ref{9}) then
determine $X^{\mu}(\tau,\sigma)$ at arbitrary value of the evolution parameter.
For different choices of initial velocities
${\dot X}^{\mu}(0,\sigma)$, ${\dot X}^{' \mu}(0,\sigma)$ we
obtain different $X^{\mu}(\tau,\sigma)$, $X^{' \mu}(\tau,\sigma)$.
In particular we can
choose ${\dot X}^{' \mu}(0,\sigma)$ such that $X^{' \mu}(\tau,\sigma)$
describes
from the mathematical point of view the same manifold $V_d$ as it
is represented
by $X^{\mu}(\tau,\sigma)$.  But physically, $X^{\mu}(\tau,\sigma)$ and $X^{'
\mu}(\tau,\sigma)$ represent different objects:  the former membrane
is deformed
in some way, and the latter membrane is deformed in some other way.\footnote
{Again we have the analogy with a usual non-relativistic elastic string
or membrane.  It can be elastically deformed in such a way that the
mathematical manifold
$V_p$ ($p=1$ or 2) remains the same, but nevertheless a deformed object
${\cal V'}_p$,
described by $\mbox{\bf x}' (\sigma)$, is physically different from the
"original" object ${\cal V}_p$
described by $\mbox{\bf x} (\sigma)$.  Both $\mbox{\bf x}
(\sigma)$ and $\mbox{\bf x}' (\sigma)$ describe the same
mathematical manifold $V_p$, but
$\mbox{\bf x}^{'} (\sigma)$ now represents positions of an elastically deformed
string or membrane.}  This illustrates that our system is indeed a "wiggly"
membrane.

The transformations at a fixed value of $\tau$
\be X^{\mu}(\sigma) \rightarrow X^{' \mu}(\sigma)  \label{9a} \ee

\nnn we interpret as {\it active transformations}
of membrane's positions (Fig.1) in spacetime.  They transform
one membrane's configuration into another configuration which may
be elastically deformed.

In general, the transformations at arbitrary values of $\tau$ and ${\sigma}^i$
\be X^{\mu}(\tau,\sigma) \rightarrow X^{' \mu}(\tau,\sigma)  \label{10} \ee
\nnn we interpret as {\it active transformations} of membrane's motion in
spacetime.  Kinematically all possible transformations of the type (\ref{10})
(with certain restrictions concerning non-singularity and single-valuedness)
are allowed, but dynamically (as relating solutions of the equations of motion)
only a subclass is allowed.

When performing quantization by using path integral approach one calculates the
transition amplitude given by the functional integral
\be <X_2(\sigma),{\tau}_2|X_1(\sigma),{\tau}_1> = \int e^{i I[X^{\mu}]} \,
     {\cal D} X^{\mu} (\tau, \sigma)  \label{11} \ee

Different functions $X^{\mu}(\tau,\sigma)$ over which the functional
integration
is performed are understood in {\it the active sense} (as described above).
They
represent various kinematically possible motions of the elastically deformed
membrane.  Since all $X^{\mu}(\tau,\sigma)$ are physically distinguishable, we
do not need to introduce ghosts.

On the contrary, in the local gauge theory of the usual $p$-branes, a class of
functions $X^{\mu}({\xi}^a)$ which can be transformed one into the other by
coordinate transformations of the worldsheet coordinates ${\xi}^a$ cannot be
interpreted as representing physically different membrane's motions.  Therefore
one needs to cancel the unphysical degrees of freedom, and a convenient way to
do this is to take into account ghosts in order to treat functional integrals
consistently.

Some more details about quantized unconstrained $p$-branes we discuss in Sec.4.
Here let us just mention that a $p$-brane's state can be represented by a wave
functional $\psi[\tau,X^{\mu}(\sigma)]$ which evolves along the evolution
parameter $\tau$.  A wave functional is in general a wave packet localized
around a "centroid" $p$-brane (Fig. 2).
As in the case of unconstrained point particle \cite{8}-\cite{9a}, a
wave packet is localized in space-time and the region of localization proceeds
foward along a time-like direction while the evolution parameter $\tau$
increases.  During such a motion in space-time the centroid
$p$-brane describes a $p+1$ dimensional worldsheet.

The wave functional is normalized in {\it space-time} so that at any $\tau$ we
have
\be \int {\psi}^{*} [\tau, X^{\mu} (\sigma)] \, {\psi} [\tau, X^{\mu} (\sigma)]
  \, {\cal D} X(\sigma ) = 1  \label{12} \ee

\nnn and consequently the evolution operator $U$ which brings $\psi (\tau)
\rightarrow \psi ({\tau}') = U\, \psi (\tau)$ is {\it unitary}. In the
particular case of a 0-brane (i.e. a point particle) the above expression
(\ref{12}) reads $\int {\psi}^{*} (\tau,x) \psi (\tau,x) {\dd}^4 x = 1$,
but in a generic case $(p \ge 1)$ the measure ${\dd}^4 x $ is replaced
by
\be
    {\cal D} X(\sigma ) = \prod_{\sigma, \mu} {\dd} X^{\mu} (\sigma)
    {\bar \gamma}^{1/4}
\label{13}
\ee

\nnn where ${\bar \gamma}$ is the determinant of the induced metric on
${\cal V}_p$.
(For details about the reparametrization invariant measure in curved
space and the origin of ${\bar \gamma}^{1/4}$ see Ref. \cite{20}.)

\vspace{1cm}

{\bf 3. Application to the embedding model of induced gravity:
a spacetime sheet
generated by a 3-brane motion}

\vspace{.3cm}

The ideas that we have developped so far may be used to describe elementary
particles as extended objects - unconstrained $p$-branes
${\cal V}_p$ - living in spacetime.
In the following we are going to follow yet another application of $p$-branes:
to represent spacetime itself!  Spacetime is considered as a surface
-called also
{\it spacetime sheet} $V_n$ embedded in a higher dimensional space $V_N$.  For
details about this model see Refs. \cite{4}-\cite{6a}.
In this section we
consider a particular model in which an $n-1$ dimensional surface, called
{\it simultaneity surface} ${\cal V}_{\Sigma}$ moves in the embedding
space according to
the uncostrained theory of Sec. 2 and sweeps an
$n$-dimensional spacetime sheet $V_n$.

Since we are now talking about spacetime which is conventionally
parametrized by
coordinates $x^{\mu}$, the notation of Sec. 2 is not appropriate. For this
particular application of the $p$-brane theory we use different notation.
Coordinates denoting position of a spacetime sheet $V_n$ (alias worldsheet) in
the embedding space $V_N$ are
\be {\eta}^a \; , \quad \quad a = 0,1,2,...,N-1  \label{13a} \ee
whilst parameters denoting positions of points on $V_n$ are
\be x^{\mu} \; , \quad \quad\mu = 0,1,2,...,n-1  \label{14} \ee
The parametric
equation of a spacetime sheet is\footnote
{To simplify notation we use the same
symbol ${\eta}^a$ to denote coordinates of an arbitrary point in $V_N$ and also
to to denote the embedding variables (which are functions of $x^{\mu}$).}
\be {\eta}^a = {\eta}^a (x)  \label{15} \ee

\nnn Parameters on a simultaneity surface ${\cal V}_{\Sigma}$ are
\be {\sigma}^i \; , \quad \quad i =1,2,...,n-1  \label{16} \ee
and its parametric equation is ${\eta}^a = {\eta}^a (\sigma)$.
In particular, one may choose such a
parametrization of $V_n$ that $x^i = {\sigma}^i$ . A moving
${\cal V}_{\Sigma}$ is described by the variables ${\eta}^a (\tau,\sigma)$.

The formal theory goes along the similar lines as in Sec. 2.  The action is
given by
\be
  I[{\eta}^a(\tau,\sigma)] = {1 \over 2} \int \omega \, {\mbox{\rm d}} \tau \,
  {\mbox{\rm d}}^{n-1}
	\sigma \sqrt{\bar f} \left ( {{ {\dot \eta}^{\mu} \,
	{\dot
	\eta}_{\mu}} \over {\Lambda}} + \Lambda \right )
\label{17}
\ee

\nnn where ${\bar f} \equiv \mbox{\rm det} {\bar f_{ij}}$, ${\bar f}_{ij}
\equiv {{\partial {\eta}^a} \over {\partial {\sigma}^i}} {{\partial {\eta}_a}
\over
{\partial
{\sigma}^j}}$ is the determinant of the induced metric on
${\cal V}_{\Sigma}$, and
$\Lambda = \Lambda(\tau,\sigma)$ a fixed function.  The tension $\kappa$ is now
replaced by the symbol $\omega$.  The latter may be a constant.
However, in the proposed
embedding model of spacetime \cite{6},\cite{6a}
we admit $\omega$ to be a function of position in
$V_N$ :  \be \omega = \omega (\eta)  \label{18} \ee In the case when
$\omega$ is a constant we have a spacetime without "matter" sources.  When
$\omega$ is a function of ${\eta}^a$, we have in general
a spacetime with sources (see Ref. \cite{6a} and Sec. 4.3)

A solution to the equations of motion derived from (\ref{17}) represents a
motion of a simultaneity surface ${\cal V}_{\Sigma}$.  This is analogous
to motion of a
$p$-brane discussed in Sec. 2.  Here again we see a big advantage
of such an {\it
unconstrained} theory:  it predicts actual {\it motion} of
${\cal V}_{\Sigma}$ and {\it
evolution} of a corresponding quantum state with $\tau$ being the evolution
parameter or historical time \cite{8a}.  The latter is a distinct concept from
the
coordinate time $t \equiv x^0$.  The existence (and progression) of a time
slice
is automatically incorporated in our unconstrained theory.  It not need be
separately postulated, as it is in the usual, constrained
relativistic theory.\footnote
{More or less explicit assumption of the existence of a time slice
(associated with the perception of "now") is manifest in conventional
relativistic theories from the very fact that the talk is about
"point-particles" or "strings" which are objects in three dimensions.}
And, since an observer cannot perceive the whole spacetime at once,
one cannot simply dispense with the existence of a time slice.
What an observer directly experiences or is aware of, are events on
${\cal V}_{\Sigma}$.  He has only
(fading) memories of the past events and expectation of future events, but he
doesn't experience neither past nor future events.  Later we shall see that
even
the concept of "time slice" is provisory and can be replaced by a suitably
generalized concept.

The theory based on the action (\ref{17}) is satisfactory in several respects.
However, it still cannot be considered as a complete theory, because it is not
manifestly invariant with respect to general coordinate transformations of
spacetime coordinates (which include Lorentz transformations).  In the next
section we shall "improve" the theory and explore some of its consequences.  We
shall see that the theory of motion of a time slice
${\cal V}_{\Sigma}$, based on the
action (\ref{17}), comes out as a particular case (solution) of the generalized
theory which is fully relativistic, i.e.  invariant with respect to
reparametrizations of $x^{\mu}$.  Yet it incorporates the concept of state
evolution.

\vspace{1cm}

{\bf 4. Spacetime as a moving 4-dimensional membrane in $V_N$ }

{\bf 4.1. General considerations}

\vspace{.3cm}

Experimental basis\footnote
{The crucial is the fact that different observers
(in relative motion) determine different sets of events as being simultaneous
and thus those events must exist in a 4-dimensional spacetime in which time is
just one of the coordinates.}
on which rests the special relativity and its
generalization to curved spacetime clearly indicates that spacetime is a
continuum in which events are existing.  On the contrary, our subjective
experience clearly tells us that not the whole 4-dimensional\footnote
{When
convenient, in order to specify the discussion, let us specify the dimension of
spacetime and take it 4.}
spacetime, but only a 3-dimensional section of it is
accessible to our immediate experience.  How to reconciliate those seemingly
contradictory observations?

It turns out that this is naturally achieved by joining the formal theory of
membrane motion (Sec. 2) with the concept of spacetime embedded in a higher
dimensional space $V_N$ (Sec. 3).  Let us assume that the spacetime is
an unconstrained 4-dimensional membrane ${\cal V}_4$ which evolves
(or moves) in the embedding space $V_N$.
What was a membrane (or $p$-brane) in Sec. 2
is now a spacetime sheet ${\cal V}_4$.  In other words,
${\cal V}_4$ (or ${\cal V}_n$ in general) is
now analogous to ${\cal V}_p$ of Sec. 2. Positions of points on ${\cal V}_4$
at a given instant of the
evolution time $\tau$ are described by embedding variables
${\eta}^a(\tau,x^{\mu})$.  The latter now depend not only on
the spacetime sheet
parameters (coordinates) $x^{\mu}$, but also on $\tau$.  Let us at the moment
just accept such a possibility that ${\cal V}_4$ evolves, and we shall
later see how the
quantized theory brings a physical sense to such an evolution.

The action which is analogous to one of eq.(\ref{17}) (which in turn is
analogous to eq.(\ref{5}) is
\be
    I[{\eta}^a(\tau,x)] = {1 \over 2} \int\omega {\mbox{\rm d}} \tau
    {\mbox{\rm d}}^4 x \sqrt{|f|} \left ( {{ {\dot \eta}^{\mu}
    {\dot \eta}_{\mu}} \over {\Lambda}} + \Lambda \right )
    \label{19}
\ee
\be
    f \equiv \mbox{\rm det} f_{\mu \nu} \; \; , \quad f_{\mu \nu} \equiv
    {\partial}_{\mu} {\eta}^a {\partial}_{\nu} {\eta}_a  \label{19aa}
\ee

\nnn where $\Lambda = \Lambda(\tau,x)$ is a fixed function of $\tau$ and
$x^{\mu}$
(like a "background field") and $\omega = \omega (\eta)$.

The action (\ref{19}) is invariant with respect to arbitrary transformations of
spacetime coordinates $x^{\mu}$.  But it is not invariant under
reparametrizations of the evolution parameter $\tau$.  Again we use analogous
reasoning as in Sec. 2.  Namely, the freedom of choice of parametrization on a
given initial ${\cal V}_4$ is trivial and it does not impose
any constraints among the
dynamical variables ${\eta}^a$ which depend also on $\tau$.  In other words, we
consider spacetime $V_4$ as a physical continuum, the points of which can be
identified and their $\tau$-evolution in the embedding space $V_N$ followed.
For a chosen parametrization $x^{\mu}$ of the points on ${\cal V}_4$ different
functions ${\eta}^a (x)$, ${\eta '}^a (x)$
(at arbitrary $\tau$) represent different {\it
physically deformed} spacetime continua ${\cal V}_4$, ${\cal V}_4^{'}$.
Different functions
${\eta}^a (x)$, ${\eta '}^a (x)$,
even if denoting positions on the same mathematical surface
$V_4$ will be interpreted as describing physically distinct spacetime
continua, ${\cal V}_4$, ${\cal V}_4^{'}$,
locally deformed in different ways.  An evolving physical spacetime continuum
${\cal V}_4$ is not identical concept to a mathematical surface $V_4$.
\footnote{A strict notation would then require a new symbol, for instance
${\tilde \eta}^a (x)$ for the variables of the physical
continuum ${\cal V}_4$,
to be distinguished from the embedding functions ${\eta}^a (x)$ of a
mathematical surface $V_4$.  We shall not use this distinction in notation,
since the meaning will be clear from the context.}

Let us now start developing some basic formalism.  The canonically conjugate
variables belonging to the action (\ref{19}) are
\be
  {\eta}^a (x)\; , \quad p_a (x) = {{\partial L} \over {\partial
   {\dot \eta}^a}}
   =\omega \sqrt{|f|} \; {{{\dot \eta}_a} \over {\Lambda}}
\label{20}
\ee

The Hamiltonian is
\be
    H = {1 \over 2} \int \, {\dd}^4 x \sqrt{|f|} \; {{\Lambda} \over
    {\omega}} \left ( {{p^a p_a} \over {\sqrt{|f|}}} - {\omega}^2 \right )
\label{21}
\ee

The theory can be straightforwardly quantized by considering ${\eta}^a (x) ,
\; p_a (x)$ as operators satisfying the equal $\tau$ commutation relations
\be
   [{\eta}^a (x) , \;p_b (x')] = {{\delta}^a}_b \delta (x - x')
\label{22}
\ee

\nnn In the representation in which ${\eta}^a (x)$ are diagonal the momentum
operator is given by the functional derivative
\be p_a = -\, i \, {{\delta} \over {{\delta} {\eta}^a (x)}}
\label{23}
\ee

\nnn A quantum state is represented by a wave functional $\psi[\tau,
{\eta}^a(x)]$ which depends on the evolution parameter $\tau$ and the
coordinates ${\eta}^a(x)$ of a physical spacetime sheet ${\cal V}_4$,
and satisfies the functional Schr\" odinger equation
\be
   i \, {{\partial \psi} \over {\partial \tau}} = H \psi
\label{24}
\ee

\nnn The Hamiltonian operator is given by eq.(\ref{21}) in which $p_a$
are now operators (\ref{23}).
A possible solution to eq.(\ref{24}) is a linear superposition of
states with definite momentum $p_a(x)$ which are taken as constant
functionals of ${\eta}^a (x)$, so that $\delta p_a / \delta {\eta}^a = 0$ :
\be
   \psi[\tau, \eta (x)] = \int {\cal D} p \, c(p) e^{- i H \tau} \,
   e^{i \int p_a (x) {\eta}^a (x) \, d^4 x}
\label{25}
\ee

\nnn where $H$ is given by (\ref{21}) and $p_a (x)$ are now eigenvalues of the
corresponding operators. The expectation value of an operator $A$ is
\be
    <A> = \int {\psi}^* [\tau, \eta (x)] A \psi [\tau, \eta(x)]
    {\cal D} \eta
\label{26}
\ee

The measures ${\cal D} \eta$ and ${\cal D} p$ should be invariant under
reparametrizations of $x^{\mu}$. This is achieved if we define
(following Ref. \cite{20})
\be
     {\cal D} \eta \equiv \prod_{a,x} {|f|}^{1/4} \mbox{\rm d} {\eta}^a (x)
\label{m1}
\ee
\be
     {\cal D} p \equiv \prod_{a,x} {|f|}^{-1/4} \mbox{\rm d} p_a (x)
\label{m2}
\ee

\nnn The above expressions result if we take into account the following
invariant scalar products
\be
      \int \mbox{\rm d} {\eta}^a (x) \mbox{\rm d} {\eta}_a (x)
      \sqrt{|f|} \, {\dd}^4 x
\label{m3}
\ee
\be
      \int {{\mbox{\rm d} p^a (x) \mbox{\rm d} p_a (x)} \over {|f|}}
      \sqrt{|f|} \, {\dd}^4 x
\label{m4}
\ee

\nnn so that the metrics are ${|f|}^{1/2} {\eta}_{ab} \, \delta (x - x')$
and ${|f|}^{-1/2} {\eta}_{ab} \, \delta (x - x')$ , respectively. Into
the definition of the invariant volume elements ${\cal D} \eta$ and
${\cal D} p$ then enter the square
roots of the determinants of the coresponding metric.

\vspace{1cm}

{\bf 4.2 A physically interesting solution}

\vspace{.3cm}

Let us now pay attention to eq.(\ref{25}). It defines a wave functional
packet spread over a continuum of functions ${\eta}^a (x)$. The expectation
value of ${\eta}^a (x)$ is
\be
       <{\eta}^a (x)> = {\eta}_{\mbox{\rm c}}^a (\tau,x)
\label{27}
\ee

\nnn where ${\eta}_{\mbox{\rm c}}^a (\tau,x)$ represents motion of the
{\it centroid} spacetime sheet ${\cal V}_4^{(c)}$ which is the "centre"
of the wave  functional packet. This is illustrated in Fig. 3.

In general, the theory admits an arbitrary motion
${\eta}_{\mbox{\rm c}}^a (\tau,x)$ which is a solution of the
classical equations of motion derived from the action (\ref{19}). But
in particular, a wave packet (\ref{25}) which is a solution of the
Schr\" odinger equation (\ref{24}) can be such that its centroid
spacetime sheet is either

\begin{description}
  \item {\  \rm (i)} \  at rest in the embedding space $V_N$, i.e.
  ${\dot \eta}_{\mbox{\rm c}}^a = 0$
\end{description}

\nnn or, more generally:

\begin{description}
  \item {\rm (ii)} \  it moves "within itself" so that its shape
  does not change
  with $\tau$. More precisely, at every $\tau$ and $x^{\mu}$ there exist
  a displacement $\Delta x^{\mu}$ such that
  ${\eta}_{\mbox{\rm c}}^a (\tau +
  \Delta \tau,x^{\mu}) = {\eta}_{\mbox{\rm c}}^a (\tau,x^{\mu} +
  \Delta x^{\mu})$ , which implies
  ${\dot \eta}_{\mbox{\rm c}}^a = {\partial}_{\mu} {\eta}_{\mbox{\rm c}}^a \,
  {\dot x}^{\mu}$ . Therefore ${\dot \eta}_{\mbox{\rm c}}^a$ is always
  tangent to a fixed mathematical surface $V_4$ which does
  not depend on $\tau$.
\end{description}

Now let us consider a special form of the wave packet as illustrated in
Fig. 4. Within the effective boundary $B$ a given function ${\eta}^a (x)$
is admissible with high probability, outside $B$ with low probability.
Such is, for instance, a Gaussian wave packet which, at the initial $\tau
= 0$, is given by
\be
  \psi [0, {\eta}^a (x)] = N e^{- \, \int \, d^4 x \sqrt{|f|} \,
  {{\omega} \over {\Lambda}} \left ({\eta}^a (x) -
  {\eta}_{\mbox{\rm c}}^a (x) \right )^2 {1 \over {2 \sigma (x)}}}
\label{28}
\ee

\nnn where the function $\sigma (x)$ vary with $x^{\mu}$ so that the
wave packet corresponds to Fig. 4.

Of special interest in Fig. 4. is the region $P$ around a spacelike
hypersurface $\Sigma$ on ${\cal V}_4^{(c)}$. In that region the wave functional
is much more sharply localized than in other regions (that is, at other
values of $x^{\mu}$). This means that in the neighborhood of $\Sigma$
a spacetime sheet ${\cal V}_4$ is relatively well defined. On the
contrary, in the regions that we call {\it past} or {\it future},
space-time is not so well defined, because the wave packet is spread over a
relatively large range of functions ${\eta}^a (x)$ (each representing a
possible spacetime sheet ${\cal V}_4$).

The above situation holds at a certain, let us say initial value of the
evolution parameter $\tau$. Our wave packet satisfies the Schr\" odinger
equation and is therefore subjected to evolution. The region of sharp
localization depends on $\tau$, and so it moves as $\tau$ increases.
In particular, it can move within the mathematical spacetime
surface $V_4$ which corresponds to such a "centroid" (physical) spacetime
sheet ${\cal V}_4^{(c)} = <{\cal V}_4>$
which "moves within itself" (case (ii) above). Such a solution of the
Schr\" odinger equation provides, on the one hand
the existence of a fixed spacetime
$V_4$, defined within the resolution of the wave packet (see Fig. 4), and
on the other hand the existence of a moving region $P$ in which the
wave packet is more sharply localized.
The region $P$ represents the "present" of an observer.
We assume
that an observer in principle measures the embedding positions ${\eta}^a (x)$
of the entire spacetime sheet. Every ${\eta}^a (x)$ is
in principle possible. However, in the practical situations available
to us, a possible measurement procedure is expected to be such that
only the embedding positions
${\eta}^a (x_{\Sigma}^{\mu})$ of the simultaneity surface $\Sigma$ are
measured with high precision\footnote
{Another possibility is to measure the induced metric $g_{\mu \nu}$ on
${\cal V}_4$, and measure ${\eta}^a (x)$ merely with a precision at
a cosmological scale or not measure at all.}
, whereas the embedding positions ${\eta}^a (x)$ of all other
regions of spacetime sheet are measured with low precision. As a
consequence of such a measurement a wave packet like one of Fig. 4
and Eq.(\ref{28}) is formed and it is then subjected to the unitary
$\tau$-evolution given by the covariant functional Schr\" odinger
equation (\ref{24}).

Using our theory, which is fully covariant with respect to
reparametrizations of spacetime coordinates $x^{\mu}$, we have thus
arrived in a natural way at the existence of time slice $\Sigma$
which corresponds to the "present" experience and which progresses
forward in spacetime. The theory of Sec. 3 is just a particular
case of this more general theory. This can be seen by taking the
limit $1/\sigma(x) \propto {\delta}^4 (x^{\mu} - x_{\Sigma}^{\mu})$
in the wave packet (\ref{28}). Then the integration over the
$\delta$-function gives in the exponent the expression
$\int {\dd}^3 x \sqrt{|{\bar f}|}({\eta}^a (x^i) -
{\eta}_{\mbox{\rm c}}^a (x^i))^2 /2 \sigma(x^i) \; i = 1,2,3$ so that
eq.(\ref{28}) becomes a wave functional of 3-dimensional membranes
${\eta}^a (x^i)$

So far we have taken that the region of sharp localization $P$ of a
wave functional packet is situated around a space-like surface $\Sigma$, and
so we obtained a time slice. But there is a difficulty with the
concept of "time slice" related to the fact that an observer in
practice never have the access to the experimental data on an
entire spacelike hypersurface. Since the signals travel with the
final velocity of light, there is a delay in getting information.
Therefore, the greater is a portion of a space-like hypersurface,
the longer is the delay. This imposes limits to the extent of
a space-like region within which the wave functional packet (\ref{28})
can be sharply localized. The situation in Fig. 4 is just an
idealization. A more realistic wave packet is illustrated in Fig. 5.
It can still be represented by the expression (\ref{28}) with a
suitably width function $\sigma (x)$.

A possible interpretation is that such a wave packet of Fig. 5 represents
a private wave function(al) of an observer. The region of sharp
localization is mainly within his brain. An observer has a relatively
good knowledge of his brain state
\footnote{This implies also a knowledge or sharp localization of
those outside regions of the spacetime sheet which are
coupled to our observer's brain by his sensory organs.}
at a given moment of the evolution time $\tau$, whilst the outside
spacetime is less well definite. It is important that the outside
spacetime is not completely indefinite; its definiteness is given by
the wave packet. So an "outside" or "objective" space-time is given
within the resolution of the wave packet. If we assume that the
wave packet moves according to the case (ii) (at the beginning of Sec. 4.2),
then the average
physical spacetime sheet $<{\cal V}_4> \leftrightarrow {\eta}_{\mbox{\rm c}}^a
(\tau,x)$
moves in such a way that all its points are within a mathematical
4-surface $V_4$ which remain constant in $\tau$. Positions on $V_4$
can be represented by $\tau$-independent embedding functions
${\eta}^a (x)$.

This model thus predicts:

\ (i) an objective existing outside spacetime $V_4$
without evolution in $\tau$ (such is a spacetime of the conventional
special and general relativity);

(ii) a region $P$ of spacetime which changes its position on $V_4$ while
the evolution time $\tau$ increases (this is a subjective region
situated mainly within the brain of an observer).

The division into an "objective" and "subjective" part
of spacetime is, of course, merely explanatory. It
serves to explain the fact that the single basic object, the
wave functional packet which moves in $\tau$, has two qualitatively
different features, as described above.

\vspace{1cm}

{\bf 4.3. Inclusion of sources}

\vspace{.3cm}

In a previous publication \cite{6a} we included the point-particle
sources into the
embedding model of gravity (which was based on the usual constrained
membrane theory). This was achieved by including in the action for
a spacetime sheet a function $\omega (\eta)$ which consits of a
constant part and a $\delta$-function part. In the analogous way we
can introduce sources into our unconstrained embedding model which
has explicit $\tau$-evolution.

For $\omega$ we can choose the folowing function of the embeding
space coordinates ${\eta}^a$ :
\be
     \omega (\eta ) = {\omega}_0 + \sum_i \int m_i {\delta}^N
     (\eta - {\hat \eta}_i ) \sqrt{|{\hat f}} \, {\dd}^m {\hat x}
\label{29}
\ee

\nnn where ${\eta}^a =  {\hat \eta}_i^a (\hat x)$ is the
parametric equation of an $m$-dimensional surface
${\hat V}_m^{(i)}$ , called {\it matter sheet}, also embedded in
$V_N \; ; {\hat x}^{\hat \mu}$ are parameters (coordinates) on
${\hat V}_m^{(i)}$ and ${\hat f}$ is the determinant of the induced
metric tensor on ${\hat V}_m^{(i)}$ . If we take $m = N - 4 +1$,
then the intersection of $V_4$ and ${\hat V}_m^{(i)}$ can be a
(one-dimensional) line, i.e. a worldline $C_i$ on $V_4$ . If $V_4$
moves in $V_N$, then also the intersection $C_i$ moves. A moving
spacetime sheet was denoted by ${\cal V}_4$ and described by
$\tau$-dependent coordinate functions ${\eta}^a (\tau,x^{\mu})$.
Let a moving worldline be denoted ${\cal C}_i$. It can be described
either by the coordinate functions ${\eta}^a (\tau, u)$ in the
embedding space $V_N$ or by the coordinate functions
$X^{\mu} (\tau, u)$ in the moving spacetime sheet ${\cal V}_4$ .
Besides the evolution parameter $\tau$ we have also a one
dimensional worldline parameter $u$ which has the analogous
role as the spacetime sheet parameters $x^{\mu}$ in
${\eta}^a (\tau ,x^{\mu})$. At a fixed $\tau$ , $\; X^{\mu} (\tau, u)$
gives a one dimensional worldline $X^{\mu} (u)$. If $\tau$
increases monotonically, then the worldlines continuously change or
{\it move}. In the expression (\ref{29}) $m-1$ coordinates
${\hat x}^{\hat \mu}$ can be integrated out and we obtain
\be
     \omega = {\omega}_0 + \sum_i \int m_i {{{\delta}^4 (x - X_i)}
     \over {\sqrt{|f|}}} \left ( {{{\dd} X_i^{\mu}} \over {{\dd}{u}}}
     {{{\dd} X_i^{\nu}} \over {{\dd}{u}}} \, f_{\mu \nu} \right )
     ^{1/2} {\dd} u
\label{30}
\ee

\nnn where $x^{\mu} = X_i^{\mu} (\tau,u)$ is the parametric equation
of a ($\tau$-dependent) worldline ${\cal C}_i$ , $u$ an
arbitrary parameter on ${\cal C}_i \, , \; \quad f_{\mu \nu} \equiv
{\partial }_{\mu} {\eta}^a {\partial}_{\nu} {\eta}_a $ the induced metric
on ${\cal V}_4$ and $f \equiv \mbox{\rm det} f_{\mu \nu}$. By
inserting Eq.(\ref{30}) into the membrane's action (\ref{19}) we obtain
the following action
   $$I[X^{\mu} (\tau, u)] =  I_0 + I_m
   = {{{\omega}_0} \over 2}  \int {\mbox{\rm d}} \tau
  {\mbox{\rm d}}^4 x \sqrt{|f|} \left ({{ {\dot \eta}^{\mu}
  {\dot \eta}_{\mu}} \over {\Lambda}} + \Lambda \right )$$
\be   + \int {\mbox{\rm d}} \tau {\mbox{\rm d}}^4 x \sqrt{|f|} \sum_i
  \left (
  {{{\dot X}_i^{\mu} {\dot X}_i^{\nu} f_{\mu \nu}} \over {\Lambda}} +
  \Lambda \right ) {{{\delta}^4 (x - X_i)} \over {\sqrt{|f|}}}
  {\left ( {{{\dd} X_i^{\mu}} \over {{\dd}{u}}}
  {{{\dd} X_i^{\nu}} \over {{\dd} {u}}} \, f_{\mu \nu} \right ) }^{1/2}
  {\dd} \lambda
\label{31}
\ee

\nnn In a special case when the membrane ${\cal V}_4$ is static with
respect to the evolution in $\tau$, i.e. all $\tau$ derivatives are
zero, then we obtain the usual Dirac-Nambu-Goto 4-dimensional membrane
coupled to point particle sources
\be
   I[X^{\mu} (u)] = {\omega}_0 \int {\dd}^4 x \, \Lambda \, \sqrt{|f|} +
   \int {\dd} u \sum_i m_i
   \left ( {{{\dd} X_i^{\mu}} \over {{\dd}{u}}}
   {{{\dd} X_i^{\nu}} \over {{\dd} {u}}} \, f_{\mu \nu} \right )
 ^{1/2}
 \label{32}
 \ee

 \nnn However, the action (\ref{31}) is more general than (\ref{32}) and
 it allows for solutions which evolve in $\tau$. The first part
 $I_0$ describes a 4-dimensional membrane which evolves in $\tau$,
 whilst the second part $I_m$ describes a system of (1-dimensional)
 worldlines which evolve in $\tau$. After performing the integration
 over $x^{\mu}$, the "matter" term $I_m$ becomes - in the case of
 one particle - analogous to the membrane's term $I_0$ :
 \be
     I_m = \int {\dd} \tau {\dd} u
     \left ( {{{\dd} X^{\mu}} \over {{\dd}{u}}}
     {{{\dd} X^{\nu}} \over {{\dd} {u}}} \, f_{\mu \nu} \right )
     ^{1/2}
     \left (
    {{{\dot X}^{\mu} {\dot X}^{\nu} f_{\mu \nu}} \over {\Lambda}} +
     \Lambda \right )
\label{33}
\ee

\nnn Instead of 4 parameters (coordinates) $x^{\mu}$ we have in (\ref{33})
a single parameter $u$, instead of the variables ${\eta}^a (\tau,x^{\mu})$
we have $X^{\mu} (\tau,u)$, and instead of the determinant of the
4-dimensional induced metric $f_{\mu \nu} \equiv {\partial}_{\mu}
{\eta}^a {\partial}_{\nu} {\eta}_a$ we have
$({\dd} X^{\mu}/{\dd} u) ({\dd} X_{\mu}/{\dd} u)$.
All what we said about the theory of an unconstrained $n$-dimensional
membrane evolving in $\tau$ can be straightforwardly applied to a
worldline (which is a special membrane with $n=1$).

After inserting the matter function $\omega (\eta )$ of eq.(\ref{29}) into
the Hamiltonian (\ref{21}) we obtain
\be
    H = {1 \over 2} \int {\dd}^4 x \sqrt{|f|} {{\Lambda} \over {{\omega}_0}}
    \left ( {{p_0^a (x) {p_0}_a (x)} \over {|f|}} - {\omega}_0^2 \right )
    + {1 \over 2} \sum_i {\dd} u
     \left ( {{{\dd} X^{\mu}} \over {{\dd} u}}
     {{{\dd} X_{\mu}} \over {{\dd} u}} \right )
     ^{1/2} {{\Lambda} \over {m_i}} \left ( P_{\mu}^{(i)} P^{(i) \mu}
     - m_i^2 \right )
\label{34}
\ee

\nnn where ${p_0}_a = {\omega}_0 {\Lambda}^{-1} \sqrt{|f|} \, {\dot \eta}_a$ is
the membrane's momentum everywhere except on the intersections
$V_4 \cap {\hat V}_m^{(i)}$, and $P_{\mu}^{(i)} = m_i {\dot X}_{\mu}/
{\Lambda}$ is the membrane momentum on the intersections
$V_4 \cap {\hat V}_m^{(i)}$. In other words, ${P_{\mu}}^{(i)}$ is the
momentum of a worldline ${\cal C}_i$ . The contribution of the
wordlines is thus explicitly separated out in the Hamiltonian (\ref{34}).

In the quantized theory a membrane's state is represented by a wave
functional which satisfies the Schr\" odinger equation (\ref{24}).
A wave packet (e.g. one of eq.(\ref{28})) contains, in the case of
$\omega (\eta)$ given by Eq.(\ref{29}), a separate contribution of
the membrane's portion outside and on the intersection
$V_4 \cap {\hat V}_m^{(i)}$ :
   $$\psi [0,\eta(x)] = {\psi}_0 [0,\eta(x)] {\psi}_m [0,X(u)]$$
\be
     = N e^{- {\omega}_0 \, \int \, d^4 x \sqrt{|f|}
  {1 \over {\Lambda}} {\left ( {\eta}^a (x) -
  {\eta}_{\mbox{\rm c}}^a (x) \right ) }^2 {1 \over {2 \sigma (x)}}}
  e^{- \sum_i \int d u {{m_i} \over {\Lambda}}
  {\left ( {{d X_i^{\mu}} \over {d u}}
  {{d X_i^{\nu}} \over {d u}} \, f_{\mu \nu} \right ) }^{1/2}
  {\left ( X_i^{\mu} (u) - X_{iC}^{\mu} (u) \right ) }^2
  {1 \over {2 {\sigma}_i (u)}}}
\label{35}
\ee
In the second factor of Eq.(\ref{35}) the wave packets of worldlines are
expressed explicitely. For a particular $\sigma (x)$, such that a wave
packet has the form as sketched in Fig. 4 or Fig. 5, there exists a region
$P$ of parameters $x^{\mu}$ at which membrane ${\cal V}_4$ is much more
sharply localized then outside $P$. The same is true for the intersections
(which are wordlines): any such a worldline ${\cal C}_i$  is much more
sharply localized in a certain interval of the worldline parameter
$u$. With the passage of the evolution time $\tau$ the region of
sharp localization on a worldline moves in space-time $V_4$. In a suitable
limit\footnote
{An analogous limit is given in Sec. 4.2 where a moving (unconstrained)
3-dimensional membrane was obtained as a limiting case of a moving wave
packet functional of a 4-dimensional membrane with a region $P$ of
sharp localization.}
this becomes equivalent to motion of a wave packet of a point particle
(event) localized in space-time. The latter particle
is just an unconstrained point particle,
a particular case (for $p=0$) of a generic unconstrained $p$-dimensional
membrane described in Sec. 2.

The unconstrained theory of point particles has a long history. It was
considered by Fock, Stueckelberg, Schwinger, Feynman, Horwitz, Fanchi,
Enatsu, and many others \cite{8}-\cite{9a}.
Quantization of the theory appeared under various
names, for instance the Schwinger proper time method or the parametrized
relativistic quantum theory. The name unconstrained theory is
used in Ref. \cite{9}
both for the classical and the quantized theory.

Such an unconstrained point particle theory implies that a wave
packet is in general
localized in space-time and the region of localization moves in space-time.
What exists is a point particle like event in space-time, not a worldline.
Then there is a problem of how to obtain the physically observed Maxwell
fields which are such that they can only be generated by charged worldline
sources, and not by charged events in space-time. A single charged event
generates the electromagnetic potential field proportional to
$e\, \delta [(x - X)^2] {\dot X}^{\mu}$, whereas a charged worldline
generates the field given by the latter expression integrated over the
parameter $\tau$ (which is usually assumed to be identical
with the worldline parameter).

While it is not quite clear whether the unconstrained point particle theory
can consistently deal with the observed electromagnetic fields\footnote
{For a discussion of the problem see Ref.\cite{8a}.}
there is no such a problem in the theory of an unconstrained
space-time membrane ${\cal V}_4$
which, for the action (\ref{31}), contains worldlines $X^{\mu} (\tau,
u)$. The latter, in addition to being existing objects in space-time
(parameter $u$), also move in space-time (parameter $\tau$). A
wave packet (\ref{35}), with $\sigma (u) $ corresponding to Fig. 4,
is expected to give the Maxwell field containing (i) a $\tau$-independent term
(as is usually observed) and (ii) a $\tau$-dependent term due to the
moving region $P$ of sharp localization. A complete treatment of the
electromagnetic fields as solutions of the dynamical equations is beyond the
scope of this paper and will be given elsewhere.

\vspace{1cm}

{\bf 5. Conclusion}

\vspace{.3cm}

We have formulated a reparametrization invariant and Lorentz invariant theory
of $p$-dimensional membranes without constraints among the dynamical variables.
This is possible if we assume a generalized form of the Dirac-Nambu-Goto
action, such that the dependence of the dynamical variables on an extra
parameter, the evolution time $\tau$, is admitted.

Such a membrane's theory manifests its full power in the embedding model
of gravity, in which space-time is treated as a 4-dimensional unconstrained
membrane, evolving in an $N$-dimensional embedding space. The embedding
model was previously discussed within the conventional theory of
constrained membranes. Release of the constraints and introduction of
the $\tau$ evolution brings new insight into the quantization of the model.
A particularly interesting is a state, represented by a functional of
4-dimensional membranes ${\cal V}_4$ , localized around an average space-time
membrane ${\cal V}_4^{(c)}$ , and even more sharply localized around a
space-like surface $\Sigma$ on ${\cal V}_4^{(c)}$. Such a state incorporates
the existence of a classical space-time continuum and the evolution. The
notorious problem of time is thus resolved in our approach to quantum
gravity. The space-time coordinate $x^0 = t$ is not time\footnote
{In this sentence "time" stands for the parameter of evolution. Such is
the meaning of the word "time" adopted by the authors who discuss
the problem of
time in general relativity. What they want to say is essentially just that
there is a big problem, since the coordinate $x^0$ cannot have the role
of evolution parameter (or "time" in short). In our work, following Horwitz
\cite{8a},
we make explicite distinction between the coordinate $x^0$ and the
parameter of evolution $\tau$. These two distinct concepts are usually
mixed and given the same name "time". In order to distinguish them, we use
the names "c{\it oordinate time}" and "{\it evolution time}".}
at all! Time must
be separately introduced, and this was achieved in our theory in which the
action depends on the evolution time $\tau$.

The importance of the evolution time was considered, in the case
of a point particle,
by many authors \cite{8}-\cite{9a}. But a charged event
in space-time generates an electromagnetic field which does not agree with
the experimentally observed field. The latter requires a worldline as a source.
Worldlines occur in our embedding model as intersections of space-time
membranes ${\cal V}_4$ with $(N-4+1)$-dimensional "matter" sheets. In the
quantized theory the state of a worldline can be represented by a wave
functional ${\psi}_m [\tau,X^{\mu}(u)]$ ,
which may be localized around an average worldline (in the
quantum mechanical sense of the expectation value). Moreover, at a certain
value $u = u_P$ of the worldline parameter the wave functional
may be much more sharply localized than at other values of $u$,
thus approximately imitating the wave function of a point particle
(or event) localized in space-time.  And since ${\psi}_m [\tau,X^{\mu}(u)]$
evolves with $\tau$, also the point $u_P$ changes with $\tau$.

The embedding model, based on the theory of unconstrained
membranes satisfying the action (\ref{19}), appears to be a promising
candidate for the theoretical formulation of quantum gravity including
the bosonic sources. Incorporation of fermions is expected to be
achieved by taking into account the Grassmann coordinates.

\vspace{1.5cm}

{\bf ACKNOWLEDGEMENT}

\vspace{.3cm}

This work was supported by the Slovenian Ministry if Science and
Technology under Contract P1-5043-0106-93.

\newpage

Fig. 1.  To different sets of initial velocities ${\dot X}^{\mu} (0,\sigma),
\; \; {\dot X }^{' \mu} (0,\sigma)$ belong different configurations
$X^{\mu} (\tau, \sigma), \; \; X^{' \mu} (\tau, \sigma)$ of membrane's
motion. They may lie on the same mathematical manifold $V_d$.

\vspace{.5cm}

Fig. 2.  A $p$-brane's wave functional is in general a wave packet localized
around a "centroid" $p$-brane. Its position in space-time depends on
the Lorentz-invariant evolution parameter $\tau$.

\vspace{.5cm}

Fig. 3.  Quantum mechanically a state of our space-time membrane
${\cal V}_4$ is given by a wave packet which is a functional of ${\eta}^a (x)$.
Its "centre" ${\eta}_{\mbox{\rm c}}^a (\tau,x)$ is the expectation value
$<{\eta}^a (x)>$ and moves according to the classical equations of motion
(as derived from the action (\ref{19}).

\vspace{.5cm}

Fig. 4.  The wave packet representing a quantum state of a space-time membrane
is localized within an effetcive boundary $B$. The form of the latter may be
such that the localization is significantly sharper around a space-like surface
$\Sigma$.

\vspace{.5cm}

Fig. 5.  Illustration of a wave packet with a region of sharp localization $P$.

\end{document}